\documentclass[twocolumn,amsmath,amssymb,nofootinbib]{revtex4}
\usepackage[english]{babel}
\usepackage[dvips]{graphicx}

\begin{document}
\title{Gamma-Rays from Dark Matter Mini-Spikes in M31}
\author{Mattia Fornasa$^1$}
\author{Marco Taoso$^1$}
\author{Gianfranco Bertone$^{1,2}$}
\affiliation{$^1$ INFN, Sezione di Padova, via Marzolo 8, Padova, 35131,
Italy} \affiliation{$^2$ Institut d Astrophysique de Paris, UMR 7095-CNRS,
Universit\'e Pierre et Marie Curie, 98bis boulevard Arago, 75014 Paris, France }
\email{mfornasa@pd.infn.it,taoso@pd.infn.it,gianfranco.bertone@pd.infn.it}

\begin{abstract}
The existence of a population of wandering Intermediate Mass
Black Holes (IMBHs) is a generic prediction of scenarios that seek
to explain the formation of Supermassive Black Holes in terms of
growth from massive seeds. The growth of IMBHs may lead to the formation of
DM overdensities called "mini-spikes", recently proposed as ideal targets for
indirect DM searches. Current ground-based gamma-ray experiments, however,
cannot search for these objects due to their limited field of view, and
it might be challenging to discriminate mini-spikes in the Milky Way
from the many astrophysical sources that GLAST is expected to observe.
We show here that gamma-ray experiments can effectively search for
IMBHs in the nearby Andromeda galaxy (also known as M31), where
mini-spikes would appear as a distribution of point-sources, isotropically
distributed in a $\thickapprox 3^{\circ}$ circle around the galactic center.
For a neutralino-like DM candidate with a mass $m_{\chi}= \mbox{ 150
GeV}$, up to 20 sources would be detected with GLAST (at $5
\sigma$, in 2 months). With Air Cherenkov Telescopes
such as MAGIC and VERITAS, up to 10 sources might be detected,
provided that the mass of neutralino is in the TeV range or above.

\end{abstract}

\maketitle

\section{Introduction}
\label{sec:chapter one}

The nature of Dark Matter (DM) is, more than 70 years after its
discovery, still an open problem. It is commonly assumed that DM is
made of Weakly Interacting Massive Particles (WIMPs), arising in
theories beyond the Standard Model (see Refs.~\cite{Bergstrom,Bertone_Hooper_Silk} for
recent reviews), the most widely discussed DM candidates being the
supersymmetric neutralino and the
lightest Kaluza-Klein particle (LKP) in theories with Unified
Extra-Dimensions \cite{Appelquist_Cheng_Dobrescu,Servant_Tait,Cheng_Matchev_Schmaltz}.
These particles will be actively searched for in upcoming high
energy physics experiments such as the Large Hadron Collider
(LHC, see e.g. Refs.\cite{Cheng_Matchev_Schmaltz,LHC site,Baltz_Battaglia_Peskin_Wizanski,Datta_Kong_Matchev} for recent discussions in the context of DM searches),
while hints on the nature of DM may already come from direct
detection experiments aiming at detecting the nuclear
recoils due to DM scattering off nuclei in large detectors (see e.g.
\cite{Munoz} for a recent update on the status of direct searches).
Alternatively, one could search for DM \emph{indirectly}, i.e. through
the detection of its annihilation products such as photons,
neutrinos, positrons and antiprotons.
The annihilation rate being proportional to the square of the DM density,
ideal targets of indirect searches include all those regions
where the DM density is strongly enhanced, due to gravitational clustering,
as in the case of the Galactic center \cite{Aharonian:2006wh,Zaharijas:2006qb,Profumo:2005xd,Mambrini:2005vk,Bertone:2005hw,Cesarini:2003nr,Bouquet:1989sr} and halo substructures \cite{Silk:1992bh,Bergstrom:1998zs,Calcaneo-Roldan:2000yt,Aloisio:2002yq, Koushiappas:2003bn, Diemand:2005vz,Pieri:2005pg,Pieri:2003cq,Baltz:2006sv}, or
because of energy losses capture in large celestial bodies, as in the case
of the Sun and the Earth (see Ref.~\cite{Bertone_Hooper_Silk} and references therein).

Large DM overdensities can also form as a consequence of astrophysical
processes, such as the adiabatic growth of Supermassive~\cite{Gondolo_Silk,Bertone:2002je,Gondolo:2000pn} or
Intermediate Mass Black Holes~\cite{Zhao:2005zr,Bertone_Zentner_Silk}.
In fact, DM halos inevitably react to the
growth of black holes, leading, in the case of adiabatic growth,
to the formation of large DM overdensities called {\it spikes}
\cite{Gondolo_Silk}. A DM cusp with a power-law density profile $\rho\propto r^{-\gamma}$,
gets redistributed after the BH growth into a steeper profile
$\rho_{sp}\propto r^{-\gamma_{sp}}$, with $\gamma_{sp} = (9 - 2\gamma) /
(4- \gamma)$, within the radius of gravitational influence of the Black Hole
(BH) (see below for further details).
BHs can thus be thought as {\it annihilation boosters}, because
the annihilation rate after their growth is boosted by several
orders of magnitude, making these objects ideal targets for indirect
DM searches.
Even in absence of mergers~\cite{Merrit_Milosavljevic_Verde_Jimenez},
and ignoring a possible off-center formation~\cite{Ullio_Zhao_Kamionkowski},
a spike around the Supermassive BH at the Galactic center
would inevitably be destroyed by the combined
effect of gravitational scattering off the {\it observed} stellar cusp
at the GC, and DM annihilations~\cite{Bertone:2005hw}. The very
same gravitational processes can still lead to the formation
of moderate enhancements called {\it crests} (Collisionally REgenerated STructures), but these
structures do not lead to significant enhancements of the annihilation signal
~\cite{Merritt:2006mt}.
{\it Mini-spikes} around Intermediate Mass Black Holes (IMBHs) are more promising targets of indirect
detection, since they would not be affected by these dynamical processes, and they
should appear as bright point-like
sources, which could be easily detected by large field of view gamma-ray
experiments as GLAST \cite{GLAST site} and further studied with ground-based Air  Cherenkov
telescopes (ACTs) \cite{Bertone_Zentner_Silk} such as CANGAROO \cite{CANGAROO site}, HESS \cite{HESS site}, MAGIC \cite{MAGIC site} and VERITAS \cite{VERITAS site}. \\
Here, we further explore the mini-spikes scenario,
and focus on the population of IMBHs in the Andromeda Galaxy (also known as
M31), a spiral galaxy very similar to the Milky Way (MW), whose center
is located $784 \mbox{ kpc}$ away from us. We compute
gamma-ray fluxes from DM annihilations around IMBHs in M31,
and show that the prospects for detection with GLAST are very
promising: in an optimistic case (a neutralino with a
mass $m_{\chi}=150 \mbox{ GeV}$ and annihilation cross section
$\sigma v = 3\times 10^{-26} \mbox{ cm}^{3} \mbox{s}^{-1}$), GLAST may
detect up to 20 point-like sources (at $5 \sigma$ and with a 2
months exposure), within $3^{\circ}$ from the center of Andromeda.
The proposed observational strategy appears particularly suited for
ACTs like MAGIC and VERITAS,
(M31 is in a region of the sky not accessible to HESS), since the
main difficulty in the search for Galactic mini-spikes
is that they cannot perform {\it deep} full-sky searches, due to their
limited field of view. In the case of mini-spikes in M31, ACTs can
search for them by scanning a small region of $\approx 3^{\circ}$ around
the center of M31, and an effective exposure of $\approx 100$ hours
in this region would be sufficient to probe the proposed scenario,
at least for DM mass in the TeV range. The next-generation Cherenkov
Telescopes Array (CTA)\cite{CTA}, is expected to significantly
improve the sensitivity, increase the field of view and decrease
the energy threshold with respect to existing ACTs, thus representing
an ideal experiment for the proposed scenario.

The paper is organized as follows: next section (Section
\ref{sec:chapter two}) is devoted to describe the formation scenario
of IMBHs. We then (Section \ref{sec:chapter three}) present the
IMBHs  catalogue and how it is adapted to the Andromeda Galaxy. We
compute the gamma-ray flux emitted by each point-like spike around
an IMBH, considering a particular energy annihilation spectrum for a
DM particle. In section \ref{sec:chapter four} we estimate the
prospects for detection for a generic ACT. Then in section
\ref{sec:chapter five} we turn to GLAST. Finally our results are
discussed in Section \ref{sec:chapter six}.

\section{Intermediate Mass Black Holes}
\label{sec:chapter two}
\subsection{IMBHs formation scenario}
IMBHs are compact objects with mass larger than $\approx 20 M_{\odot} $,
the heaviest remnant of a stellar collapse \cite{Fryer_Kalogera},
and smaller than $ \approx 10^{6} M_{\odot} $, the lower end of
the mass range of SuperMassive Black
Holes (SMBH) \cite{Ferrarese_Ford}. The theoretical and observational
motivations for IMBHs were recently reviewed in Ref.~\cite{Miller_Colbert}. For instance,
Ultra-Luminous X-ray point sources (ULXs) could be interpreted
as accreting IMBHs, since alternative explanations in terms of AGNs,
neutron stars or SMBHs appear to be problematic or even ruled
out~\cite{Miller_Colbert,Swartz_Ghosh_Tennant_Wu}.

From a theoretical point of view, a population of massive seed
black holes could help to explain the origin of SMBHs.
In fact, observations of quasars at
redshift $z\approx 6$ in the
Sloan Digital survey ~\cite{Fan:2001ff,barth:2003,Willott:2003xf}
suggest that SMBHs were already in place
when the Universe was only $\sim 1$ Gyr old, a circumstance
that can be understood in terms of rapid growth starting
from massive seeds (see e.g. Ref.~\cite{haiman:2001}).

In fact, a generic prediction of scenarios that seek to explain the
properties of the observed SuperMassive Black Hole population,
is that a large number of ``wandering''
IMBHs should exist in DM halos~\cite{islama:2003,volonteri:2003,Koushiappas_Bullock_Dekel}.
Despite their theoretical interest, it is difficult to obtain
conclusive evidence for the existence of IMBHs.
A viable detection strategy could be the search
for gravitational waves produced in the mergers of the IMBH
population~\cite{Thorne:1976,Flanagan:1998a,Flanagan:1998b,islamd:2004,
Matsubayashi:2004,Koushiappas:2005qz}, with space-based
interferometers such as LISA~\cite{LISA site}.

In Ref.~\cite{Bertone_Zentner_Silk}, two scenarios for IMBHs
formation have been considered. The first posits IMBHs as remnants of the
collapse of Population III stars. Zero-metallicity Population
III stars are more massive than more recent metal-enriched stars and, if
heavier than $ 260 M_{\odot} $, they would collapse directly into black
holes (\cite{Miller_Colbert} and Refs. therein).

Here, we will focus only on the second scenario, based on
Ref.~\cite{Koushiappas_Bullock_Dekel}, where IMBHs form
at high redshift from gas collapsing in mini-halos. If the latter
are massive enough, proto-galactic disks form at the center of each
halo. Gravitational instabilities introduce an effective viscosity
that causes an inward mass and an outward angular momentum flow. The
process goes on till it is interrupted by feedback from star
formation (1-10 Myrs) that heats the disk. Then the so-formed object
undergoes gravitational collapse into a black hole. A characteristic
mass scale of $ 10^7 M_{\odot} $ is imprinted to the mini-halo by
the requirements that it is heavy enough to form a gravitational
unstable disc and that the black hole formation timescale is shorter
than the  typical major mergers one.
The resulting black holes have a mass log-normally scattered, with a
$ \sigma_{\bullet} = 0.9 $, around the mean value of~\cite{Koushiappas_Bullock_Dekel}:
\begin{eqnarray}
M_{\bullet} & = & 3.8 \times 10^4 M_{\odot} \left( \frac{\kappa}{0.5} \right)
\left( \frac{f}{0.03} \right)^{3/2} \left( \frac{M_{vir}}{10^7 M_{\odot}}
\right)  \\
& & \times \left( \frac{1+z}{18} \right)^{3/2} \left(
\frac{t}{\mbox{ 10 Myr}} \right) \nonumber,
\end{eqnarray}
where $ \kappa $ is that fraction of the baryonic mass which loses
its angular moment that remains in the remnant black hole. $ f $ is
the fraction of the total baryonic mass in the halo that has fallen
into the disc, $ M_{vir} $ is the halo virial mass, $ z $ is the
redshift of formation and $ t $ the timescale for the evolution of
the first generation of stars.\\

In Ref.~\cite{Bertone_Zentner_Silk} Bertone et al. have studied the
population of IMBHs in the MW, following the evolution of mini-halos
hosting IMBHs at high redshift (populated with the prescriptions of
Ref.~\cite{Koushiappas_Bullock_Dekel}), down to redshift $z=0$ (see
Ref.~\cite{Bertone_Zentner_Silk} for further details). As a result,
they obtained 200 realizations of the IMBHs population in the
Galaxy, that were used to produce 200 mock catalogs of DM
mini-spikes, and to study the prospects for detection of these
objects in the Galaxy. The average number of unmerged IMBHs was
found to be $ \overline{N} = 101 \pm 22 $, and each of these objects
is characterized by its mass, distance from the center of the
galaxy, and surrounding DM distribution.

\subsection{DM distribution around IMBHs}

Following earlier work on the dynamics of stars and DM around compact
objects (see Ref.~\cite{Quinlan_Hernquist_Sigurdsson} and references therein),
Gondolo and Silk have shown that the adiabatic growth of a massive
black hole in the center of a dark halo modifies the distribution of
the surrounding DM, inducing an enhancement of the density called
"spike" \cite{Gondolo_Silk}. They focused their attention on the SMBH
at the center of our Galaxy, but the same formalism can be applied also to
IMBHs. The initial DM distribution in all mini-halos can be
adequately parametrized  with a Navarro, Frenk and White (NFW)
profile~\cite{Navarro_Frenk_White}:
\begin{equation}
\rho(r)=\rho_0\left(\frac{r_s}{r}\right)\left(1+\frac{r}{r_s}\right)^{-2},
\label{eqn:NFW}
\end{equation}
where $ r_s $, called the scale radius, sets the radius at which the
profile slope changes. The new profile after the adiabatic growth,
will be \cite{Gondolo_Silk}:
\begin{equation}
\rho_{sp}(r)=\rho(r_{sp})\left(\frac{r}{r_{sp}}\right)^{-7/3},
\label{eqn:powerlaw}
\end{equation}
where $ \rho $ is the density function of the initial NFW profile.
$r_{sp} $ gives the upper limit inside which Eq. \ref{eqn:powerlaw}
is considered valid and is related to the radius of gravitational
influence of the black hole $r_{h}$: $r_{sp} \thickapprox 0.2 r_{h}$
\cite{Merritt}, where $r_h$ is implicitly defined as:
$$ M(r<r_h) \equiv \int_{0}^{r_{h}} \rho(r)r^{2}dr = 2 M_{\bullet} $$
with $M_{\bullet}$ is the mass of the black hole.\\

The spike profile diverges at low radii but annihilations set
an upper limit to the physical density. Solving the evolution
of the DM particles number density, one finds that the upper limit depends on the
microphysical properties of the DM particles (mass and annihilation cross
section) and on the evolution timescale of the black hole. We denote the distance where
the density equals this upper limit $ r_{lim} $, and following
Ref.~\cite{Bertone_Zentner_Silk} we define a cut-radius for our density profiles:
\begin{equation}
r_{cut}=\mbox{Max}[4R_{Schw},r_{lim}],
\end{equation}
where $ R_{Schw} = 2.95$ km $M/M_{\odot}$ is the BH Schwarzschild radius.
The density between $ R_{Schw} $ and $ r_{cut} $ is
assumed to be constant to $ \rho_{sp}(r_{cut}) $.
\begin{figure}[t]
\includegraphics[width=8.5cm]{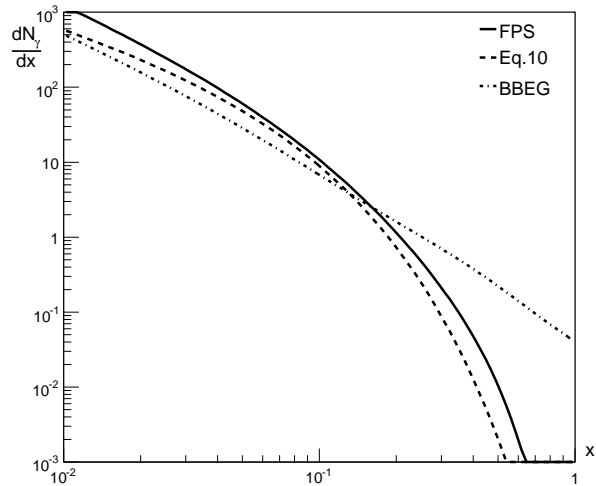}
\caption{Differential photon spectrum per annihilation. Different
parametrizations and annihilation channels are shown. Solid line
(FPS) is an analytic fit relative to the $b\bar{b}$ channel,
as obtained in Eq. \ref{eqn:FPS}. Dashed line
(Eq.~\ref{eqn:BSS}) is relative to the same annihilation channel
$b\bar{b}$, but with a different parametrization of the FFs
(see Eqs.~\ref{eqn:fx BSS} and \ref{eqn:BSS}). Dotted line
(BBEG) is relative to $B^{1}$ annihilations and includes final
state radiation from annihilation to charged leptons \cite{Bergstrom_Bringmann_Eriksson_Gustafsson}
(see text for more details)}
\label{fig:frammentazioni}
\end{figure}
\section{Gamma-Rays from IMBHs in M31}
\label{sec:chapter three}

\subsection{IMBHs in M31}
\begin{figure*}[t]
\includegraphics[width=8.5cm]{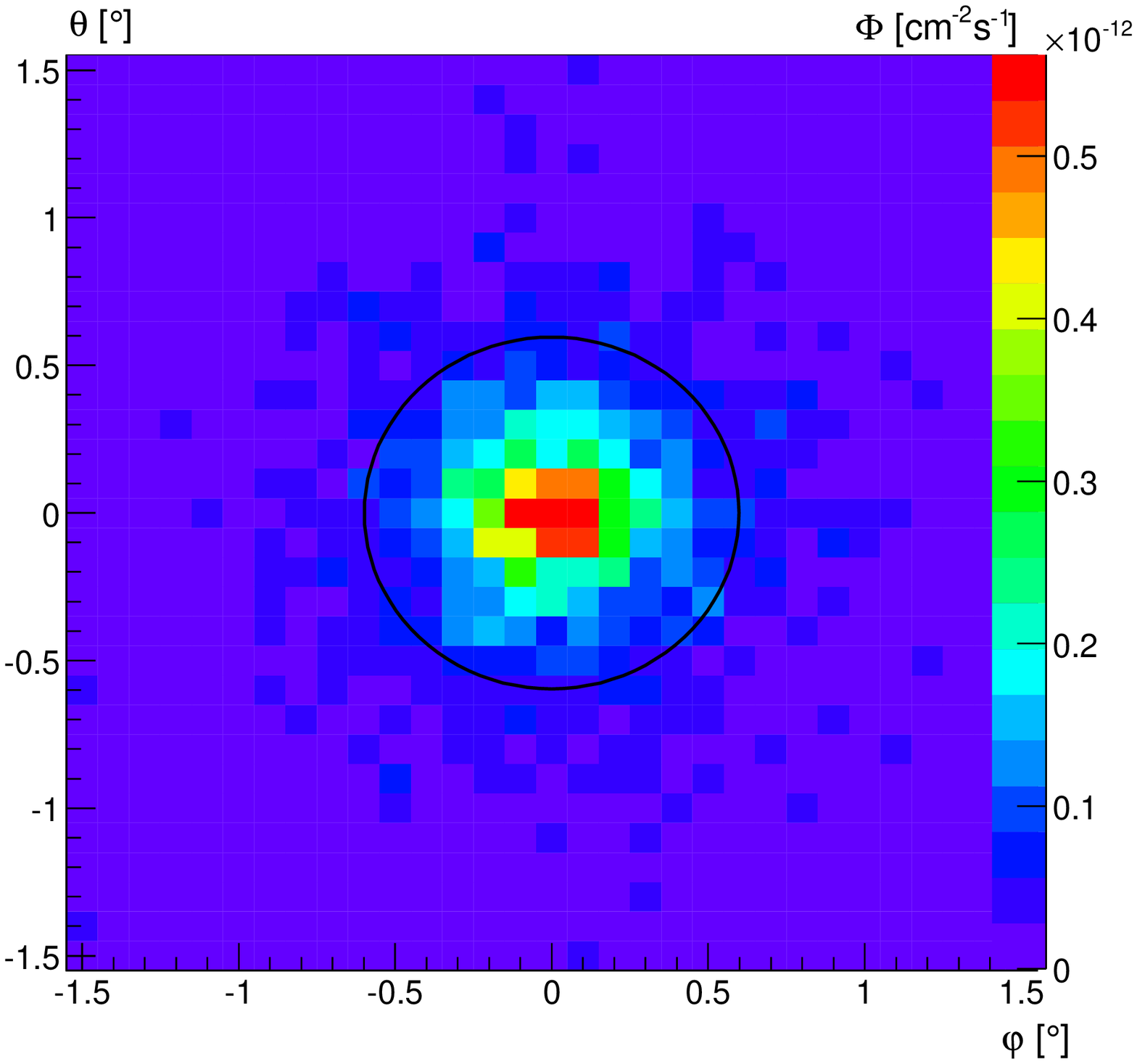}
\includegraphics[width=8.5cm]{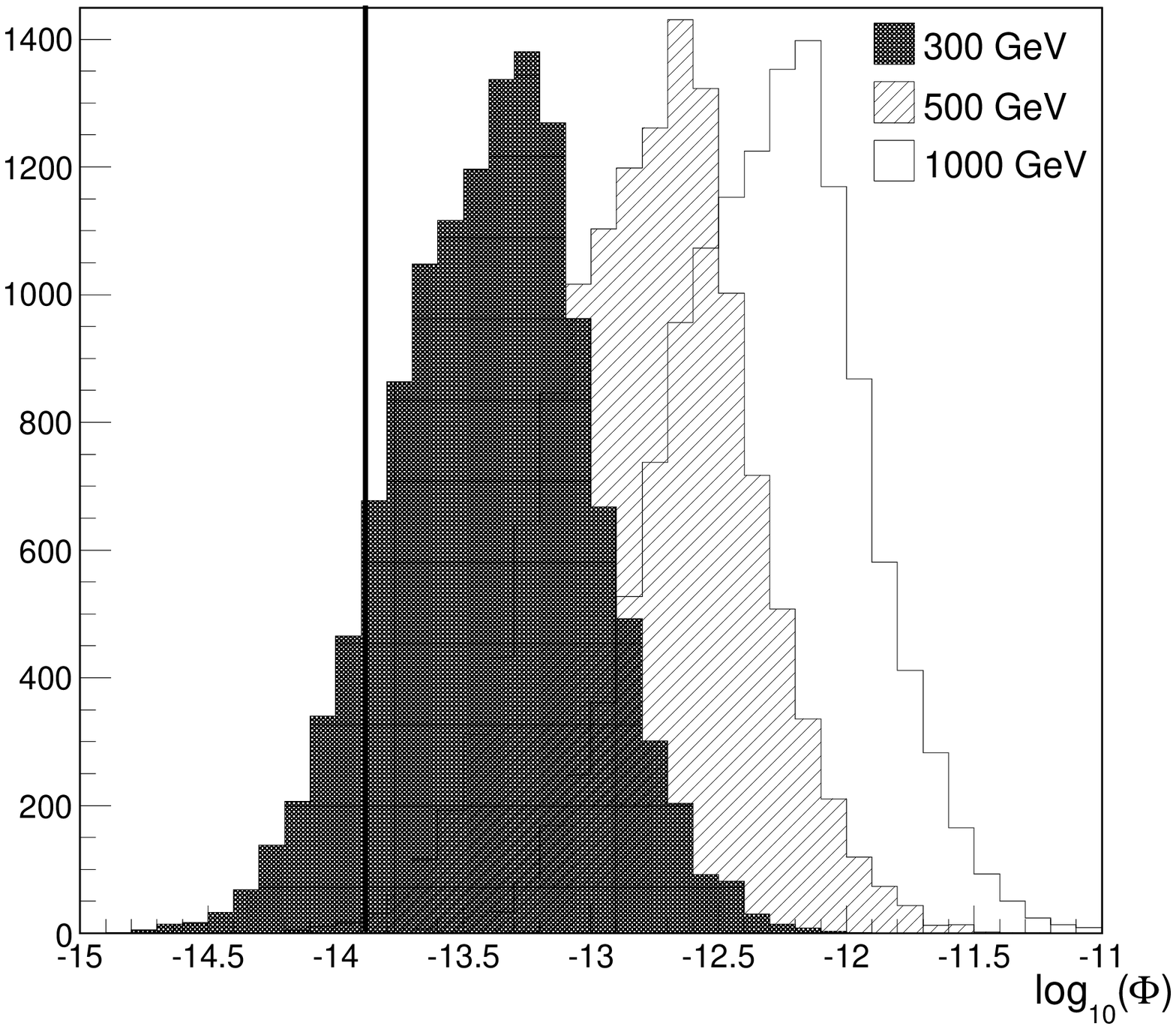}
\caption{{\it Left:} Map of gamma-rays emission (above 100 GeV, in
$\mbox{cm}^{-2}\mbox{s}^{-1} $) by DM annihilation ($m_\chi =1$ TeV) around IMBHs
in Andromeda, averaged over all realizations. Bins are $
0.1^{\circ} $ wide, to match the angular resolution of  ACTs
and GLAST. The circle shows for comparison the M31 scale radius $ r_s $.
{\it Right:} Luminosity function of IMBHs (fluxes are in $ \mbox{cm}^{-2}\mbox{s}^{-1}$),
for  $ m_{\chi}=0.3, 0.5$ and 1 TeV. The vertical line shows the contribution of the
smooth component of the M31 halo, assuming a NFW profile and $m_{\chi} = 1 \mbox{ TeV}$.}
\label{fig:averagemap}
\end{figure*}

Although similar, the Milky Way and Andromeda do not have exactly the same
properties. The mock catalogs of IMBHs built for our Galaxy, thus have to be
modified to account for the different average number and different spatial
distribution in the host halo. A comparison between the properties of
Andromeda and of the Galaxy is shown in Table
\ref{tab:Andromeda characteristics}.

\begin{table}[b]
\begin{tabular}{|c|c|c|}
\hline
 & Milky Way & Andromeda \\
\hline
Distance to the center [kpc] & 8.5 & 784.0 \\
Virial Radius [kpc] & 205 & 180 \\
Virial Mass [$ M_{\odot} $]& $ 1.0 \times 10^{12} $ & $ 6.8 \times 10^{11} $ \\
$ r_s $ [kpc] & 21.75 & 8.18 \\
$ \rho_0 $ [$ \frac{M_{\odot}}{kpc^3} $] & $ 5.376 \times 10^6 $ & $ 3.780 \times 10^7 $\\
\hline
\end{tabular}
\caption{Distance from the Sun (in kpc), virial radius (defined as the radius within which the density reaches 200 times the critical density, in kpc), virial mass (in solar masses) and the two  NFW density profile parameters (in kpc and $ M_{\odot} \mbox{kpc}^{-3} $ respectively), both for the MW and the Andromeda Galaxy \cite{Fornengo_Pieri_Scopel,Geehan_Fardal_Babul_Guhathakurta}.}
\label{tab:Andromeda characteristics}
\end{table}
We start from the mock catalogs obtained in Ref.~\cite{Bertone_Zentner_Silk}
and we rescale the total number of objects by the ratio between the host halo masses,
since the number of unmerged IMBHs scales linearly
with the host halo mass, and the galactocentric distance by the ratio
of virial radii. We obtain for M31 an average number of IMBHs per realization
$N_{M31}= 65.2 \pm 14.5$.
The mass spectrum remains unchanged, with an average mass around
$10^5 \mbox{ } M_{\odot} $, while the average distance from
the center of the galaxy is $ 32.31 $ kpc. We have verified that our rescaling
procedure reproduces in a satisfactory way the properties of the IMBHs
population in Andromeda, by comparing our results with a limited number of
mock catalogs obtained as an exploratory study in Ref.~\cite{Bertone_Zentner_Silk}.

\subsection{Gamma-rays flux from IMBHs in M31}
Once the mock catalogs of IMBHs in M31 have been obtained, it is possible to
calculate the gamma-ray flux from each IMBH in every realization.
The calculation follows Eq. 14 in Ref. \cite{Bertone_Zentner_Silk}:
\begin{eqnarray}
\label{eqn:flux}
\Phi(E) & = & \frac{\sigma v}{2m^2_{\chi}}\frac{1}{d^2} \frac{dN_{\gamma}(E)}{dE} \int_{r_{cut}}^{r_{sp}}\rho^2(r)r^2 dr \\
& = & \Phi_0 \frac{dN_{\gamma}(E)}{dE} \left( \frac{\sigma
v}{10^{-26}\mbox{cm}^3/\mbox{s}} \right) \left( \frac{m_{\chi}}{1
\mbox{ TeV}} \right)^{-2}
\nonumber \\
& & \times \left( \frac{d}{\mbox{780 kpc}} \right)^{-2} \left( \frac{\rho(r_{sp})}{100 \mbox{ GeV}/\mbox{cm}^3} \right)^2  \nonumber \\
& & \times \left( \frac{r_{sp}}{5\mbox{ pc}} \right)^{14/3} \left(
\frac{r_{cut}}{10^{-3} \mbox{ pc}} \right)^{-5/3} \nonumber,
\end{eqnarray}
where $\Phi_{0}= 2.7 \times 10^{-14} \mbox{ cm}^{-2}\mbox{s}^{-1}$,
$ d $ is the IMBH distance to the observer, $\sigma v$ is the DM
annihilation cross section times relative velocity and $ m_{\chi} $
is the DM particle mass (the letter $ \chi $, usually adopted for
neutralino, is used here to denote a generic WIMP candidate). $
r_{cut} $ and $ r_{sp} $, represent the inner and outer size of the
spike, as discussed in the previous section.

$ dN_{\gamma}(E)/dE $ is the differential photon yield per
annihilation, that can be expressed as:
\begin{equation}
\frac{dN_{\gamma}(E)}{dE}=\sum_{a}B^a \frac{dN_{\gamma}^a(E)}{dE},
\end{equation}
where $ B^a $ is the branching ratio $ \mbox{BR}(\chi\chi \rightarrow
a\bar{a}) $ and $ dN_{\gamma}^a/dE $ the secondary photon spectrum
relative to the annihilation channel $a\bar{a} $.
The latter term is thus a purely Standard
Model calculation, while branching ratios have to be derived in the framework
of new theories beyond the Standard Model, such as SUSY or UED.

We review here different parametrizations of the photon yield that
have been recently proposed in literature.
The first parametrization we focus on, has been obtained in
Ref.~\cite{Fornengo_Pieri_Scopel}, and it is relative to
annihilations into $ b\bar{b} $. The authors have parametrized the
results obtained with the event generator PYTHIA \cite{PYTHIA}
as follows
\begin{equation}
\frac{dN_{\gamma}^{b}(x)}{dx}=x^a\mbox{e}^{b+cx+dx^2+ex^3},
\label{eqn:FPS}
\end{equation}
where the parameters depend on the neutralino mass, and
for the specific case $ m_{\chi}=1 \mbox{ TeV} $,
$ (a,b,c,d,e)=(-1.5,0.37,-16.05,18.01,-19.50) $.
While for annihilation to $ \tau$s
\begin{equation}
\frac{dN_{\gamma}^{\tau}(x)}{dx}=x^a(bx+cx^2+dx^3)\mbox{e}^{ex},
\label{eqn:tau}
\end{equation}
and for $ m_{\chi}=1 \mbox{ TeV} $, $ (a,b,c,d,e)=(-1.31,6.94,-4.93,-0.51,-4.53) $.

Alternatively, one can start from the most recent
Fragmentations Functions (FFs) (e.g. \cite{Kretzer}), describing the
hadronization of partons into the particles of interest.
The FF of $ b $ quarks hadronizing in neutral pions has been fitted with a
simple analytic form that captures in a satisfactory way the behavior of the
FF at large $ x $ finding the following analytic fit
\begin{equation}
f(x)=\frac{7.53}{x^{0.87}e^{14.62x}}.
\label{eqn:fx BSS}
\end{equation}
Convolving the spectrum pions with their decay spectrum into photons
one finally obtains the differential photon yield
\begin{equation}
\frac{dN_{\gamma}(x)}{dx}=\int_{x}^{1}f(x^{\prime})\frac{2}{x^{\prime}}dx^{\prime}.
\label{eqn:BSS}
\end{equation}

We have also considered an example inspired from theories
with Unified Extra-Dimensions, where the role of DM is usually played by the
first excitation of the hypercharge gauge boson, and referred to as
$ B^{(1)} $. Since the $ B^{(1)} $ annihilation
into fermions does not suffer from chirality suppression, as
in MSSM, we also include the contribution from annihilation to
$l\bar{l}\gamma $, as calculated in Ref.~\cite{Bergstrom_Bringmann_Eriksson_Gustafsson},
as well as the contribution from $ \tau $ fragmentation, and usual
from annihilations to $ b\bar{b} $, with the appropriate branching ratio.
The final state radiation arising from annihilation to charged leptons
has a characteristic, very hard, spectral shape~\cite{Beacom:2004pe,Bergstrom_Bringmann_Eriksson_Gustafsson}
\begin{equation}
\frac{dN_{\gamma}^{l}(x)}{dx}=\sum_{l=e,\mu} \frac{\alpha}{\pi}\frac{x^2-2x+2}{x}\ln \left[ \frac{m^2_{B^{(1)}}}{m^2_l}(1-x) \right].
\label{eqn:charged leptons}
\end{equation}

The three prescriptions for the annihilation spectrum
are plotted in Fig. \ref{fig:frammentazioni} (for $m_\chi = 1$ TeV).
As expected, all spectra are very similar up to $x \equiv E / m_\chi \sim 0.1$,
but the spectrum relative to $ B^{(1)} $ annihilations is harder at large
$ x $, and exhibits a distinctive sharp cut-off at $x=1$.

To show the small effect that the adoption of different annihilation
spectra has on the prospects for indirect detection, we have calculated
the DM annihilation flux from the smooth component of the M31 halo,
assuming a NFW profile with the parameters described in
Tab. ~\ref{tab:Andromeda characteristics} above. The results are
displayed in Table \ref{tab:Andromeda flux}, and as one can see,
differences are within a factor of 2. In the remain of this
paper, we will
thus work only with the first analytic fit, since the uncertainties
associated with other astrophysical and particle physics parameters are
significantly larger.

\begin{table}[b]
\begin{tabular}{|c|c|}
\hline
  & M31 flux [$ \mbox{cm}^{-2}\mbox{s}^{-1} $] \\
\hline
FPS \cite{Fornengo_Pieri_Scopel} & $ 1.33 \times 10^{-14} $ \\
Eq.~\ref{eqn:BSS} & $ 9.79 \times 10^{-15} $ \\
BBEG \cite{Bergstrom_Bringmann_Eriksson_Gustafsson} & $ 1.60 \times 10^{-14} $ \\
\hline
\end{tabular}
\caption{Gamma-ray flux over 100 GeV from Andromeda (in $
\mbox{cm}^{-2}\mbox{s}^{-1} $) for a smooth NFW, and for the
different parametrizations discussed in the text. Differences among
the predicted fluxes are within a factor of 2.} \label{tab:Andromeda
flux}
\end{table}
\begin{figure}[t]
\includegraphics[width=8.5cm]{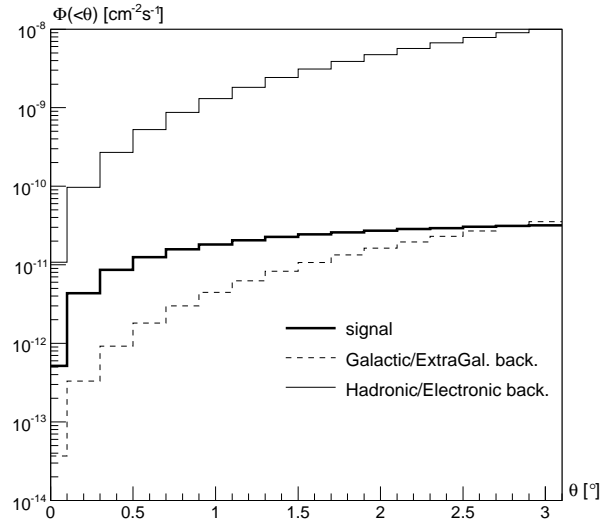}
\caption{Gamma-ray flux (in $ \mbox{cm}^{-2}\mbox{s}^{-1} $) from DM
annihilation around IMBHs (solid thick line), integrated over a cone
of size $\theta$ towards the center of M31, as a function of $\theta$.
We show for comparison the
hadronic/electron background, assuming $ \epsilon_h=10^{-2} $ (solid
thinner line) and the diffuse extragalactic background (dashed
line).} \label{fig:background}
\end{figure}

By calculating the gamma-ray flux in Eq.~\ref{eqn:flux} for IMBHs in
all realizations, we obtain a gamma-ray map, above $E_{thr}=100$
GeV, from mini-spikes in Andromeda. Each M31 realization actually
produces a different emission map, so in  Fig. ~\ref{fig:averagemap}
we show the average of all 200 maps, which clearly exhibits, as
expected, a strong enhancement of the flux in the innermost regions
of the Galaxy. The pixel size matches the angular resolution of
ground-based telescopes such as VERITAS and MAGIC, and of GLAST. For
the map, a DM particle mass of 1 TeV and an annihilation cross
section $ \sigma v =3 \times 10^{-26} \mbox{cm}^{3}\mbox{s}^{-1} $
have been adopted. Note that this map alone does not provide any
information on the detectability of the fluxes, which will be
discussed in detail in the next 2 sections. Note also that the
actual distribution of IMBHs, will provide (as we shall see later) a
much more 'patchy' emission, far from the average smooth behaviour
shown in Fig.~\ref{fig:averagemap}.

We also show in the right panel of Fig.~\ref{fig:averagemap} the luminosity
function of IMBHs (sum of all realizations), for different values of the DM particle mass.
The distribution is approximately gaussian, and the average flux of IMBHs is
larger than emission due the smooth component. The dependence from
the mass results in due to a balance between the $m_\chi^{-9/7}$ dependence
in Eq.~\ref{eqn:flux}, and the $m_\chi$ dependence of the upper limit in the integral
of the energy spectrum. Having set in the figure an energy threshold
$ E_{thr} = 100$ GeV, the luminosity flux towards higher fluxes when the mass
increase. We will come back later to this threshold effect, that leads to
higher fluxes for higher masses when $m_\chi \sim E_{thr}$ despite the
explicit $m_\chi^{-9/7}$ dependence of the annihilation flux.
Meanwhile we note that this effect disappears when $m_\chi \gg  E_{thr}$,
as can be seen from Table \ref{tab:average_flux}.
\begin{table}[b]
\begin{tabular}{|c|c|}
\hline
 & Average flux [$ \mbox{cm}^{-2}\mbox{s}^{-1} $] \\
\hline
$ m_{\chi}=50 \mbox{ GeV} $ & $ 5.26 \times 10^{-11} $ \\
$ m_{\chi}=150 \mbox{ GeV} $ & $ 7.65 \times 10^{-11} $ \\
$ m_{\chi}=300 \mbox{ GeV} $ & $ 6.92 \times 10^{-11} $ \\
$ m_{\chi}=500 \mbox{ GeV} $ & $ 5.81 \times 10^{-11} $ \\
\hline
\end{tabular}
\caption{Average flux from IMBHs
in all 200 realizations (in $ \mbox{cm}^{-2}\mbox{s}^{-1} $), for
different values of DM mass, $ \sigma v=3 \times 10^{-26}
\mbox{cm}^3\mbox{s}^{-1} $ and $E_{thr} =4$ GeV.}
\label{tab:average_flux}
\end{table}
\begin{figure*}[t]
\includegraphics[width=9cm]{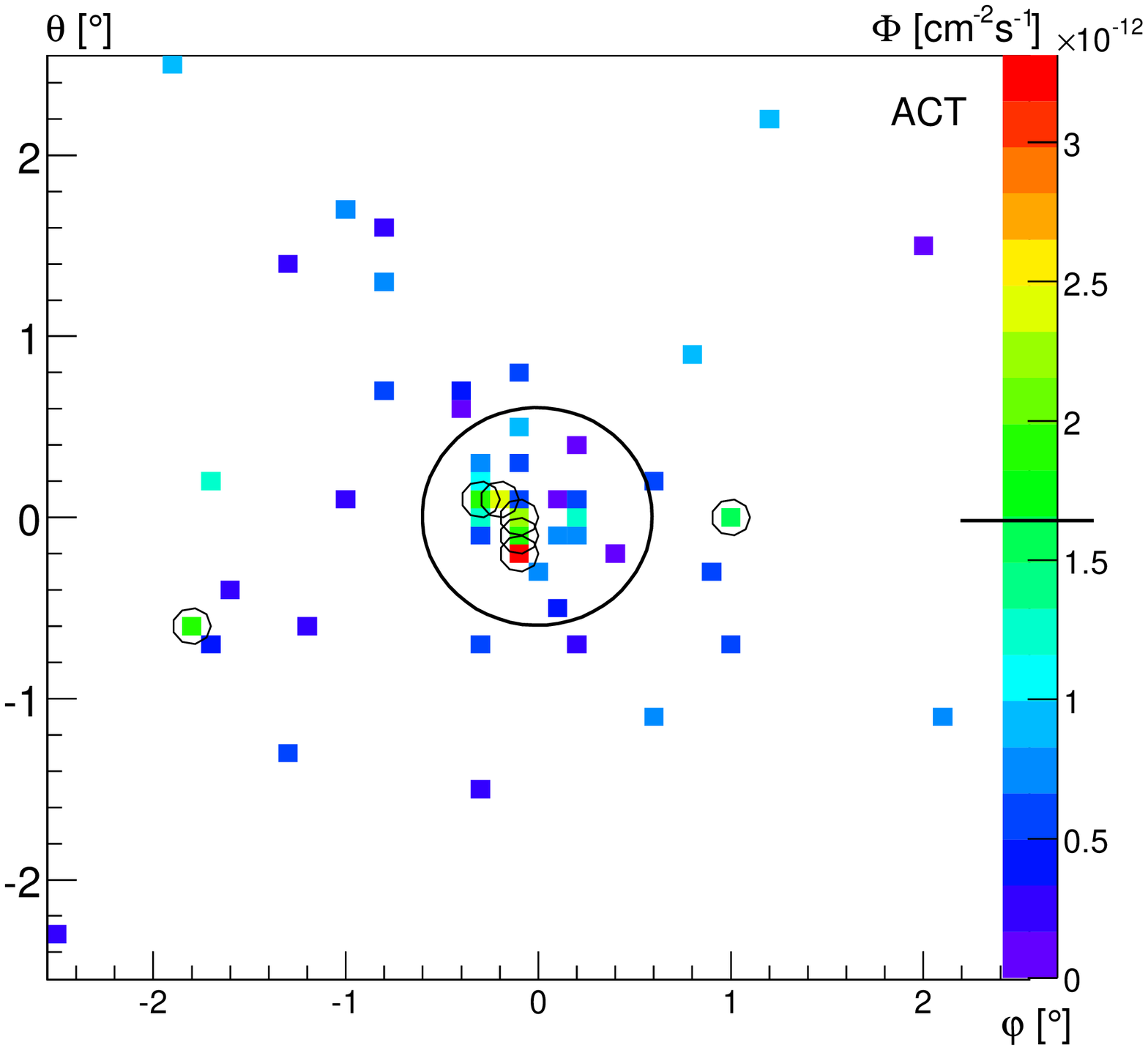}
\includegraphics[width=8.5cm]{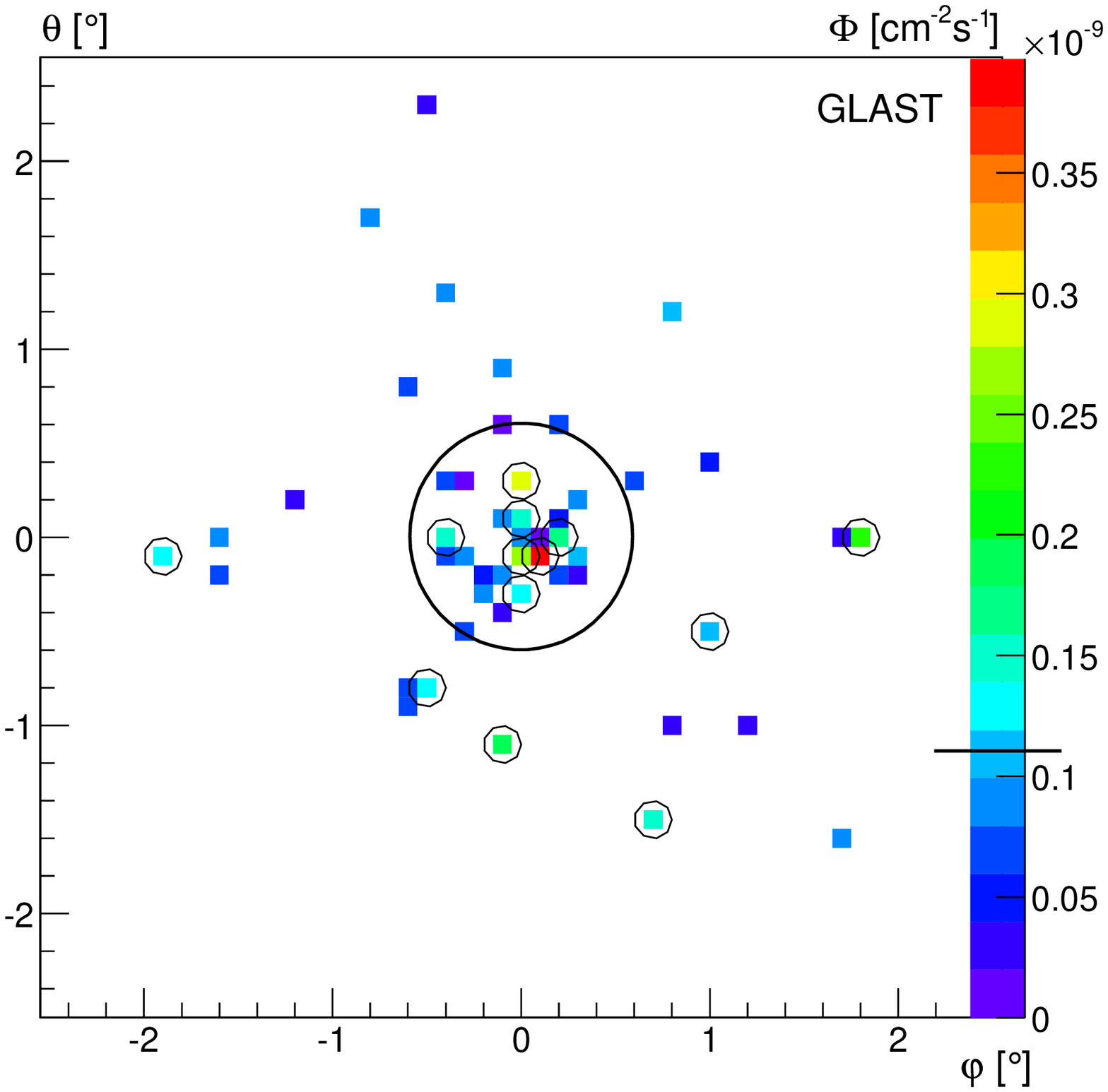}
\caption{Left (right) panel shows a map of the gamma-ray flux in
units of photons $\mbox{cm}^{-2}\mbox{s}^{-1} $, from DM
annihilations around IMBHs in M31, relative to one random realization
of IMBHs in M31. The size of the bins is $0.1^{\circ}$
and the threshold for the left (right) panel is 100 GeV (4 GeV)
as appropriate for ACTs (GLAST). The circles highlight IMBHs
within the reach of current ACTs for a $5 \sigma$ detection in
100 hours (within the reach of GLAST for a
5$\sigma$
detection in 2 months).}
\label{fig:First_MA}
\end{figure*}

\section{Prospects for detection}

As we shall see the prospects for detection depend on the expected
or measured experimental performances, but also on the atmospheric
and astrophysical backgrounds. We perform separate
analysis for Air Cherenkov Telescopes and the upcoming gamma-ray
satellite GLAST.

\subsection{Prospects for ACTs}
\label{sec:chapter four}
The calculations in this section are performed for a generic ACT,
but they are particularly relevant for two specific experiments:
MAGIC and VERITAS. As for HESS, being located in Namibia, it cannot
detect gamma-rays from the direction of Andromeda.

To determine the significance of the signal from an individual
mini-spike, as calculated in the previous section, we compare the number
of signal photons, to the fluctuations of the background
\begin{equation}
n=\frac{n_{\gamma}}{\sqrt{n_{bk}}} = \sqrt{T \cdot\Delta
\Omega}\frac{\int A_{eff}(E,\theta)\frac{d\Phi}{dE} dE
d\theta}{\sqrt{\int A_{eff}(E,\theta)\frac{d\Phi_{bk}}{dE} dE
d\theta}}, \label{eqn:sigma}
\end{equation}
where $ T $ is the
exposure time, $ A_{eff} $ the effective area,  $ \Delta\Omega $ the
solid angle, $d\Phi_{bk} /dE$ is the total
background differential flux.

For Air Cherenkov Telescopes, the main
background is due to hadrons interacting with the atmosphere and
producing electromagnetic showers. Following
Ref.~\cite{Gaisser_et_al} \cite{Pieri:2003cq}, we consider
\begin{equation}
\label{eqn:hadron}
 \frac{d\Phi_h}{d\Omega dE}=1.5 \times
\left( \frac{E}{\mbox{GeV}} \right)^{-2.74}\frac{\mbox{p}}{\mbox{cm}^2\mbox{s GeV sr}}.
\end{equation}
\begin{figure*}[t]
\includegraphics[width=8.5cm]{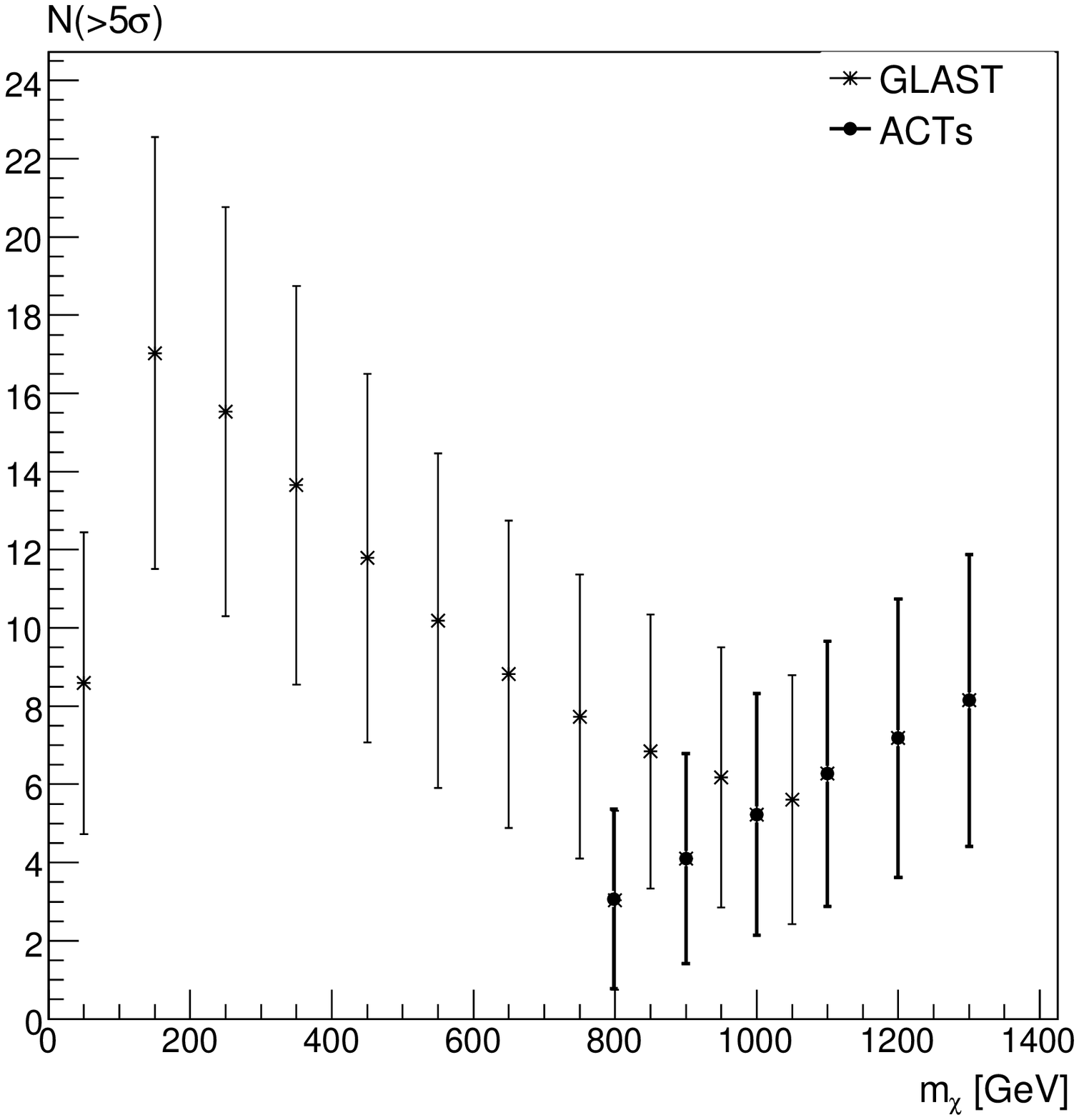}
\includegraphics[width=8.5cm]{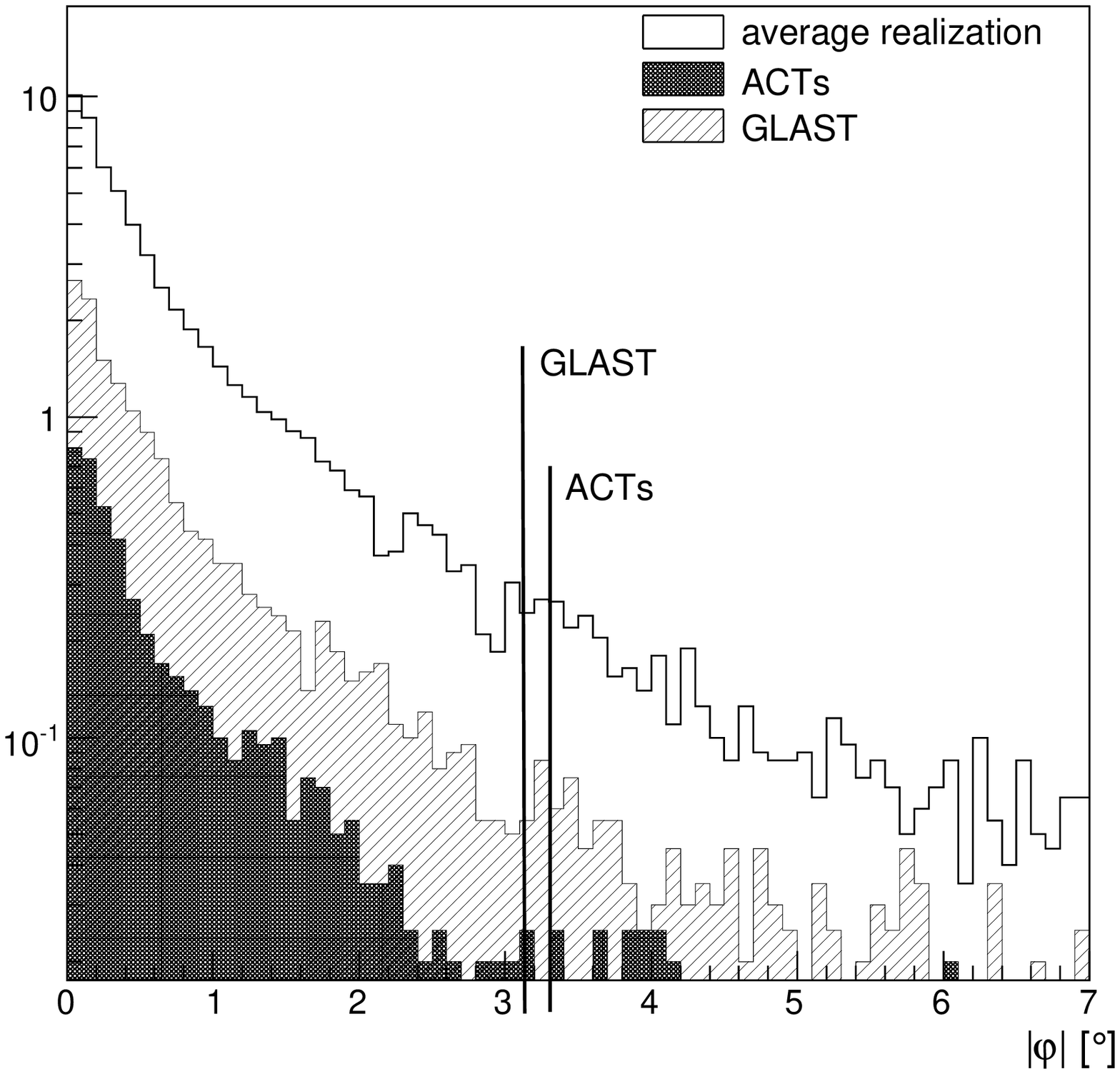}
\caption{Number of detectable mini-spikes in M31 with GLAST (2 months)
and ACTs (100 hours) as
a function of the DM particle mass (left) and as a function of the
angular distance from the center of M31 (right). In the left panel, error
bars denote the $1-\sigma$ scatter among different realizations. In the
right panel, the total number of objects is shown as an empty histogram,
while the vertical lines denote the size of the region that contains 90\% of the
detectable IMBHs.} \label{fig:Radiale}
\end{figure*}

The ratio of the number of hadrons misinterpreted as gamma-rays, over the total
number of cosmic ray hadrons, $\epsilon_{h}$,
provides an estimate of the telescope potential to discriminate the
gamma-ray signal from the hadronic background. We adopt a typical value
$ \epsilon_{h}=10^{-2} $, following Refs.~\cite{Pieri:2003cq}
\cite{Bergstrom_Hooper}. The electronic contribution to the background
is~\cite{Pieri:2003cq}:
\begin{equation}
\label{eqn:electron}
\frac{d\Phi_e}{d\Omega dE}=6.9 \times 10^{-2}
\left( \frac{E}{\mbox{GeV}} \right)^{-3.3}\frac{\mbox{e}}{\mbox{cm}^2\mbox{s GeV sr}}
\end{equation}
and it is typically subdominant at the energies of interest.

In Figure \ref{fig:background} we compare the DM annihilation signal with the
different sources of background, as a function of the field of view.
The minimum flux for a 5$ \sigma $ detection with an effective area
of $ A_{eff} = 3 \times 10^4 \mbox{ m}^2 $ \cite{MAGIC site} and an exposure
time of 100 hours, is $ \phi_{min}=1.6 \times 10^{-12} \mbox{cm}^{-2}\mbox{s}^{-1} $.

To produce this estimate we have considered values of effective area
and angular resolution similar to MAGIC and the result is consistent
with earlier estimates of the MAGIC sensitivity~\cite{Bertone_Hooper_Silk}.
An actual estimate of the instrument performance suggests that the minimum
flux can be up to an order of magnitude higher~\cite{Mose}.\\

The fluxes in Fig.~\ref{fig:averagemap} are thus found to lie below the
minimum detectable flux so determined, so one may na\"ively conclude
that there is no hope to detect them with this experimental setting.
However, for a more careful assessment of the prospects for detection,
we need to estimate the detectability in each realization and {\it then}
average over all the realization, and not viceversa. In fact, as
demonstrated in the left panel of Fig.~\ref{fig:First_MA}, in each random realization,
the emission is much more patchy, with a large number of high emission peaks,
corresponding to individual mini-spikes, that can thus be resolved with
the adopted angular resolution.
Black circles highlight the position of objects brighter than the
experimental sensitivity (indicated in the color scale by the black line).
In total, for $m_{\chi}= 1 \mbox{ TeV}$, the number of detectable IMBHs is
$N_{5\sigma}=5.2\pm3.1$, where the error is relative to the 1-$\sigma$
scatter among different realizations.

We note that current simulations indicate that the next-generation Cherenkov Telescopes
Array (CTA)\cite{CTA}, may significantly improve the sensitivity, down to
$ \phi_{min} \approx 10^{-13}\mbox{cm}^{-2}\mbox{s}^{-1}$, thus leading to
a substantial improvement in the prospects for detection.

\subsection{Prospects for GLAST}
\label{sec:chapter five}

The space satellite GLAST is expected to play a crucial role in
indirect DM searches, thanks both to its ability to perform observations
at energy scales comparable to the mass of common
DM candidates and to its potential of making deep full-sky maps in
gamma-rays, thanks to its large ($\sim 2.4$ sr) field of view~\cite{GLAST site}.
Despite the smaller effective area, it is not affected,
being a satellite, by the atmospheric hadronic and electron background.
Furthermore, its lower energy threshold (30 MeV) allows to probe
lighter DM particles, typically leading to higher fluxes.
The angular resolution of GLAST is $\approx 3^{\circ} $ in the
energy range 30 MeV-500 MeV, becomes $ 0.5^{\circ} $ from 500 MeV
to 4 GeV, and reaches $ 0.15^{\circ} $ above 4 GeV \cite{GLAST
site}.

As in the case of ACTs, we compare the expected fluxes with
the photon background, which in this case, since GLAST will perform
observations above the atmosphere, is mainly due to diffuse gamma-ray emission.
The galactic and extragalactic background has been measured in
\cite{Cillis_Hartman,Cillis_Hartman_web_page} by EGRET in the energy range between 30 MeV
and 10 GeV and we extrapolate it to higher energies by fitting with
a power-law with spectral index of -2.1. The resulting formula is
\begin{equation}
\label{eqn:extra/gal}
 \frac{d\Phi_{extra/gal}}{d\Omega dE}=2.3 \times
10^{-6}\left(\frac{E}{\mbox{GeV}}\right)^{-2.1}\frac{\gamma}{\mbox{cm}^2\mbox{s GeV sr}}.
\end{equation}

We note here that a large fraction of the observed
gamma-ray background might be actually due to DM annihilations ~\cite{Bergstrom:2001jj,Taylor:2002zd,Ullio:2002pj,Ando:2005hr}, in particular
if astrophysical processes can boost the annihilation signal~\cite{Horiuchi:2006de,Ahn:2007ty}.
In this case, the smoking-gun for this scenario would be the
distinctive shape of the angular power-spectrum of the background,
that may allow, already with GLAST, the discrimination against ordinary astrophysical
sources~\cite{Ando:2006cr}.

The sensitivity above 30 MeV, i.e. the minimum detectable flux for
a 5$ \sigma $ detection with an exposure of 2 months, is found to be
$ \phi_{min}=3.2 \times 10^{-8} \mbox{cm}^{-2}\mbox{s}^{-1} $. This
value is derived from Eq.~\ref{eqn:sigma}, adopting values of the
energy dependent effective area provided by~\cite{Rando}, and is consistent
with GLAST sensitivity maps obtained in Ref. \cite{Bertone:2006kr}.
The integral flux above threshold from IMBHs, averaged among realizations and
integrated in a $ 3^{\circ} $ cone towards M31, is
$ \phi_{30} = 1.3 \times 10^{-7}\mbox{cm}^{-2}\mbox{s}^{-1}$.

In the right panel of Fig.~\ref{fig:First_MA}, we show the results
of our analysis relative to a random realization, and adopting
$m_\chi = 150$ GeV, and $\sigma v = 3 \times 10^{-26}$ cm$^3$
s$^{-1}$. Mini-spikes appear as high emission peaks, and can be
easily resolved by selecting photons above 4 GeV, so that the
angular resolution of GLAST approaches 0.1 degrees.  Black circles
highlight those objects that produce a flux detectable at $5\sigma$
with GLAST, with a 2 months exposure.

\section{Discussion and Conclusions}
\label{sec:chapter six}

Although we have performed the analysis of the prospects for detection with GLAST
and ACTs for 2 different benchmark scenarios (essentially high DM particle mass
for ACTs, low $m_\chi$ for GLAST), the analysis can be easily extended to any
value of the particle physics parameters of the annihilating DM particle. To explore the
dependence on $m_\chi$, we show in the left panel of Fig.~\ref{fig:Radiale} the number of
objects that can be detected with the aforementioned experiments, as a function
of the DM particle mass. Near the experiment threshold,
fluxes increase with mass. When $m_\chi \gg E_{thr}$ this threshold effect disappears and
one recovers the expected behavior (smaller fluxes for higher masses).

Similarly, one can plot the number of detectable objects as a function
of the angular distance from the center of M31, to estimate the region
where most mini-spikes can be found. This is shown in the right panel of Fig.~\ref{fig:Radiale}
where the total number of objects is also shown for comparison.
Vertical lines denote the angular size of the region that contains 90\% of the
detectable IMBHs for the various experiments, which has a characteristic size of
$ \theta = 3.3^{\circ} $.

We stress that while in the case of Galactic IMBHs the identification of
mini-spikes will require a case-by-case analysis of the spectral properties
of unidentified gamma-ray sources, the detection of a cluster of sources
around the center of Andromeda would {\it per se} provide unmistakable
evidence for the proposesd scenario.

In conclusion, we have computed gamma-ray fluxes from DM annihilations in
mini-spikes around IMBHs in the Andromeda Galaxy. We have studied the prospects
for detection with
Air Cherenkov telescopes like MAGIC and VERITAS and with the GLAST satellite,
and found that a handful of sources might be within the reach of current ACTs,
while the prospects for the planned CTA are more encouraging. The obvious
advantage of the proposed scenario with respect to mini-spikes in the MW, is that they are not randomly distributed over
the sky, but they are contained, at 90\%, within 3 degrees from the center
of Andromeda, and can thus be searched for with ACTs by performing a deep
scan of this small region.

The prospects for GLAST appear more promising, since an exposure time of
2 months allows the detection of up to of $\approx 20$ mini-spikes, that would be
resolved as a cluster of point-sources with identical
spectra, within a $\sim 3^\circ$ region around the center of Andromeda.
Such a distinctive prediction cannot be mimicked by ordinary
astrophysical sources.
As in the case of IMBHs in the MW, null searches would place very strong constraints
on the proposed scenario in a wide portion of the DM parameter space.

\centerline{\bf Acknowledgements} We thank Riccardo Rando, of the
GLAST collaboration, as well as Mos\`e Mariotti and Michele Doro of
the MAGIC collaboration, for useful information and discussions on
experimental strategies and sensitivities. We also thank Andrew
Zentner for earlier collaboration and for providing the mock
catalogs of IMBHs in the MW and Lidia Pieri for comments. GB is
supported by the Helmholtz Association of National Research Centres.

\end{document}